\documentclass{article} 
\usepackage{bm}% bold math

\def\beq{\begin{equation}} 
\def\eeq{\end{equation}} 
\begin{document}
\title{The light-front vacuum and dynamics\footnote{Presented at Light-Cone 2004, Amsterdam, 16 - 20 August} \footnote{
This work supported 
in part by the U.S. Department of Energy, under contract DE-FG02-86ER40286.
}}

\author{W.N. Polyzou}
\maketitle 
\vspace{0.3cm} 
%\begin{document}
%
\maketitle
\begin{abstract}
I give a quantum theoretical description of kinematically invariant
vacuua on the algebra of free fields restricted to a light front and
discuss the relation between the light-front Hamiltonian, $P^-$, the
vacuum, and Poincar\'e invariance.  This provides a quantum theoretical
description of zero modes.

\end{abstract}

\section{Introduction}
A light-front field, $\phi (f)$, is a free field restricted to the
light front $x^+=0$ and smeared with a Schwartz test function of the
light-front variables $\tilde{x}=(x^-, \vec{x}_{\perp})$.  The
commutator of two light-front fields is:

\begin{equation}
[\phi (f), \phi (g) ]_- = {1 \over 2}[(f,g)_f - (g,f)_f]
\qquad
(f,g)_f := \int {d\tilde{p}\theta (p^+)  \over p^+} \tilde{f}^*(\tilde{p}
) \tilde{g}(\tilde{p})
\label{eq:AB}
\end{equation}
where $\tilde{f}(\tilde{p})$ is the Fourier transform of
$f(\tilde{x})$.  In the absence of restrictions on the test functions,
the light-front scalar product in (\ref{eq:AB}) diverges
logarithmically due to the $p^+=0$ singularity in the denominator.
The commutator (\ref{eq:AB}) becomes finite if the test functions are
restricted to have the form $\tilde{f}(\tilde{p}) = p^+
\tilde{g}(\tilde{p})$, where $\tilde{g}(\tilde{p})$ are ordinary
Schwartz test functions of the light-front momenta.  This space of
test functions, introduced by Schlieder and Seiler \cite{ss}, is
denoted by ${\cal S}^+$.

The light-front Fock algebra, ${\cal A}_f$, is generated by
finite linear combinations of the form
\begin{equation}
A := \sum_{k=1}^N c_k e^{i \phi (f_k)}
\label{eq:BH}
\end{equation}
where $c_k$ are complex and $f_k(\tilde{x})$ are real Schlieder-Seiler
functions.  It is straightforward to show that ${\cal A}_f$ is an
abstract $*$-algebra with following properties:
\begin{itemize}
\item [1.] ${\cal A}_f$ is closed under kinematic 
Poincar\'e transformations. 
\item [2.] ${\cal A}_f$ is a Weyl algebra.
\item [3.] Rotations induce algebraic isomorphisms from ${\cal A}_f$
to algebras associated with different light-front orientations.
\end{itemize}
A vacuum is a positive, invariant, linear functional $E[\cdot ]$ on
this algebra.  The properties of a vacuum can be expressed in terms of
its generating functional
\[
S\{ f \} := E[e^{i \phi (f)}] = 
\langle 0 \vert e^{i \phi (f) } \vert 0 \rangle.
\]
The generating functional of a light-front vacuum 
must be normalized, $S\{ 0 \}=1$,
real $S^* \{ f \}=S\{ -f^* \} $, continuous
\[
f_n \to f \in {\cal S}^+ \Rightarrow S\{f_n \} \to S\{ f \},
\]
non-negative 
\[
S\{f_i -f_j\} := M_{ij} \geq 0
\]
for any sequence $\{ f_n\}$ of real test functions in ${\cal S}^+$,
kinematically invariant 
\[
S\{ f\}= S\{f'\}
\]
where $f'(x) = f(\Lambda x+a)$ for any {\it kinematic} Poincar\'e
transformation, and satisfy cluster properties
\[
\lim_{\lambda \to \infty} S\{ f+ g_\lambda  \}\to  S\{f\}S\{ g\} 
\]
where $g_\lambda (x) = g (x + \lambda y)$
and $y$ is any space-like vector in the light-front hyperplane
\cite{fcwp}.

The Hilbert space representation of ${\cal A}_f$
associated with a given vacuum functional is defined as follows.
A dense set of vectors is given by expressions of the form 
\[
\vert \psi \rangle= \sum_{n=1}^N c_n e^{i \phi (f_n)} \vert 0 \rangle 
\qquad N< \infty .
\]
The inner product can be expressed in terms of the generating functional
\[
\langle \xi \vert \psi \rangle := 
\sum_{mn}d_m^* c_n e^{{1 \over 2} [\phi (g_m) ,\phi (f_n) ]}  
S\{f_n -g_m \} .
\]
The generating functional of the Fock representation, $S_0\{ f \}:=
e^{-{1 \over 4} (f,f)}$, satisfies all of the required properties.
The representation of the kinematic Poincar\'e transformations on this
Hilbert space is unitary.

The algebra ${\cal A}_f$ has another class of kinematically invariant
vacuua. Given a Schlieder-Seiler test function $\tilde{f}(\tilde{p})$
define
\begin{equation}
\hat{f} (\vec{p}_{\perp}) = \lim_{p^+ \to 0} {\tilde{f} 
(p^+, \vec{p}_{\perp}) \over p^+} .
\end{equation}
Vacuum generating functionals have the form
\begin{equation}
S\{ f \}= S_0\{ f \} s\{ \hat{f} \}
\qquad 
s\{ \hat{f} \} =
e^{\sum_n i^n s_n (\hat{f}, \cdots ,\hat{f})}
\label{eq:sf}
\end{equation} 
where 
\begin{equation}
s_n (\vec{p}_{1\perp},  \cdots, \vec{p}_{n\perp}):= 
\delta (\sum_{i=n} \vec{p}_{i \perp} )
w_{tn} (\vec{p}_{1 \perp}, 
\cdots, \vec{p}_{n \perp} ) 
\end{equation}
and $w_{tn} (\vec{p}_{1 \perp}, \cdots , \vec{p}_{n \perp} )$ are
connected, two-dimensional, Euclidean invariant Schwartz distribution
in $2(n-1)$ independent variables.  The functional, $s\{\hat{f}\}$, is
the Fourier transform of a positive measure on the cylinder sets of
Schwartz distributions in two variables \cite{gelf}.  The generating
function $S\{f\}$ will be the generating function of a vacuum
functional if $s\{ \hat{f} \}$ satisfies the same properties as
$S\{f\}$ with respect to Schwarz functions of two variables, where the
invariant subgroup is the two-dimensional Euclidean group.  While
positivity is a strong constraint, non-trivial examples associated
with coherent states and Gaussian measures exit.  The different vacuum
generating functionals lead to inequivalent Hilbert space
representations of the $*$ algebra.

To construct a relativistic model it is necessary to complete the 
Lie algebra of the Poincar\'e group by finding dynamical 
generators that are compatible with a given vacuum state.
The first step is to construct a mass operator that is compatible
with the vacuum.  Let $O$ be any non-negative kinematically 
invariant operator.  The Fock representation mass operator, $M_f$, 
is one such example; others can be constructed by choosing  
$O= B^{\dagger}B$ for a kinematically invariant operator $B$.
Let $\Pi:= (I- \vert 0 \rangle \langle 0 \vert)$ .

The operator $M:= \Pi O\Pi $ is non-negative, kinematically invariant, 
and annihilates the vacuum. It is a suitable candidate for the mass operator 
of a unitary representation of the Poincar\'e group.
The associated light-front Hamiltonian $P^-$ is:
\[
P^- := {\vec{P}_{\perp} \vec{P}_{\perp} + M^2 \over P^+} .
\]
The mass operator $M$ can be formally expressed as a limit of 
elements of ${\cal A}_f$.  To do this let  
$\{A_n \} \in {\cal A}_f$ generate an orthonormal basis for 
the Hilbert space representation with vacuum functional $E$,  
$E[A_n^{\dagger}A_m ] = \delta_{mn}$.  It follows that
\[
M = \sum_{k,l} A_k^{\dagger} A_l m_{kl} \qquad \mbox{where}
\]
\[
m_{kl} := E[A_k^{\dagger} O A_l ]
-E[A_k^{\dagger}] E[O A_l]
-E[A_k^{\dagger}O]E[A_l]
+ E[A_k^{\dagger}] E[ O]E[ A_l ] 
\]
which explicitly exhibits $M$ as the limit of elements of the 
algebra. 
% The vacuum dependence of the expansion coefficients 
%shows how this operator must change to remain compatible
%with the vacuum vector.

The last step in constructing a dynamics is to complete the Lie
algebra of the Poincar\'e group by including rotations.  The ability
to complete the Lie algebra is intrinsic to the choice of
kinematically invariant $M$.  Free rotations, $U_0(R)$, act on the
fields covariantly and define algebraic isomorphisms from
${\cal A}_f$ to light-front Fock Algebras with {\it different} light
fronts.  Given $M$ and the kinematic observables it is possible to
formulate a scattering theory.  The point eigenstates of the mass 
operator are needed to formulate the scattering asymptotic condition.
They are acceptable if under {\it free} rotations they satisfy 
\[
U_0 (R) \vert m \rangle_{\hat{n}} = \vert m \rangle_{R\hat{n}} D(R)
\]
where $D(R)$ is an irreducible representation of the rotation group
(note the change in the orientation of the light front).
Wave operators, constructed using these one-body solutions to 
formulate the asymptotic condition, satisfy 
\[
U_0 (R) \Omega_{\hat{n} \pm} = \Omega_{R\hat{n} \pm} U_f (R)
\]
where $\Omega_{R\hat{n} \pm}$ is the wave operator associated with the
rotated light front, $R\hat{n}$ and $U_f(R)$ is the asymptotic representation 
of $SU(2)$.  Poincar\'e invariance requires
rotation operators that leave the vacuum invariant.  If the scattering
operators are asymptotically complete and {\it independent} of the
orientation of the light-front,
\[
\Omega^{\dagger}_{+ \hat{n}} \Omega_{- \hat{n}} =
\Omega^{\dagger}_{+ R\hat{n}} \Omega_{- R\hat{n}} 
\Rightarrow 
A_R:= \Omega_{+ R\hat{n}} \Omega^{\dagger}_{+ \hat{n}}  =
\Omega_{- R\hat{n}} \Omega^{\dagger}_{- \hat{n}} 
\]
then it follows that 
\[
U_{\hat{n}}(R) := U_0(R)  A_{R^{-1}} =
 U_0 (R) \Omega_{R^{-1} \hat{n} \pm} 
\Omega^{\dagger}_{\hat{n} \pm}   =
\Omega_{\hat{n} \pm} U_f (R)  
\Omega^{\dagger}_{\hat{n} \pm}   
\]
extends the Poincar\'e group on the Hilbert space with the given
non-trivial vacuum \cite{wp}.  The invariance of the $S$ matrix ensures that
$A_R$ does not depend on the choice of asymptotic condition.  Here
wave operators are two-Hilbert space wave operators that necessarily
include the point spectrum contributions to the mass operator.  The
$\hat{n}$-invariance of $S$ can be tested in calculations,
however the restrictions are non-trivial.

The above construction shows that the singularities in the
light-front scalar product lead to a restriction on the test functions
in the light-front Fock algebra.  The resulting $*$-algebra, ${\cal A}_f$,
has a large
class of kinematically invariant vacuua that lead to inequivalent
Hilbert space representations of the algebra.
It is
possible to find dynamical $P^-$ or mass operators $M$ that are
positive, annihilate the vacuum and are kinematically covariant
(resp. invariant).  A large class of such operators exist, but a
given operator can only annihilate one of the vacuum vectors.  The
dynamical operator $P^-$ or $M$ can be used to formulate a scattering
theory.  A sufficient condition for full Poincar\'e invariance is that
the scattering matrix associated with different light fronts is
independent the light front.  This is a strong, but testable
condition.  

The construction provides a direct means to formulate dynamical models
associated with inequivalent vacuum representations.  This
construction does not utilize classical equations of motion or models
with a finite number of degrees of freedom, it provides a provide
direct means to formulate and study models with non-trivial zero
modes.

\end{document}